\newif\ifpdf
\ifx\pdfoutput\undefined
  \pdffalse              
\else
  \pdfoutput=1           
  \pdftrue
\fi

\documentclass[aps,prb,twocolumn,superscriptaddress]{revtex4}

\ifpdf
  \usepackage[pdftex,
        colorlinks=true,
        pdftitle={A New face on old code},
        pdfauthor={P.F. Peterson and Th. Proffen},
        pdfsubject={},
        pdfkeywords={NOBUGS2002/024},
        pagebackref,
        pdfpagemode=None,
        bookmarksopen=true]{hyperref}
  \usepackage{thumbpdf}
\fi
\usepackage{graphicx}
\usepackage{amssymb}

\begin{document}
\title{Little helpers for your experiment - NOBUGS2002/013}

\author{Th. Proffen}
\affiliation{Los Alamos National Laboratory, LANSCE-12, Mailstop H805, Los
Alamos, NM 87544, USA}

\author{P.F. Peterson}
\affiliation{Intense Pulsed Neutron Source, Building 360, 9700 South Cass
Ave, Argonne, IL 60439-4814, USA}

\date{October 20, 2002}


\begin{abstract}
The World Wide Web (WWW) is a wonderful tool to provide users with
documentation and tools that make the preparation and running of an
experiment a little easier. We present a set of tools that allow one to
estimate the activation of a sample in the neutron beam, or to calculate
the absorption of a sample. Ever tried to figure out the time for the next
sample change - here our scheduler can help. In addition there is a simple
database that can be used to keep track of all the data files collected at
various facilities. We present these little tools, but also demonstrate
how easy it is to create new WWW based tools that make the life of users
easier.
\end{abstract}

\maketitle

\section{Introduction}

Visiting a user facility such as a synchrotron or neutron source for the
first time to carry out an experiment can be a overwhelming experience.
One of the major stepping stones is to learn how to use the control
software for the instrument and how to extract, process and view the data
collected. But there are also minor issues such as, how do I find out
about the activation of my sample after a two day neutron diffraction
experiment. The World Wide Web (WWW) is a great tool to provide
documentation that help users to prepare for an experiment and to find
answers to common questions during the experiment. While much can be
accomplished by static WWW documents, a lot of useful help can be added by
including interactive WWW pages. A perfect example are the various
interactive pages offered to users by the Institut Laue-Langevin in
Grenoble, France at http://barns.ill.fr. Another example is a diffraction
physics tutorial \cite{prnebi01} which allows students to simulate
disordered structures and explore the corresponding scattering pattern.
This is achieved by interfacing the diffuse scattering simulation program
DISCUS \cite{prne97} with a WWW based form allowing students to change
various parameters of the simulation.
\par
In this paper we will briefly show how easy it is to add a WWW form
interface to a given program and we present three of our own examples. Here
we are mainly concerned with the WWW interface, in a separate paper
\cite{thomas;cabs037} we discuss how old programs can be improved by
adding a graphical users interface.

\section{Simple example}

\begin{figure}[!tb]
  \centering \includegraphics[angle=0,width=2.8in]{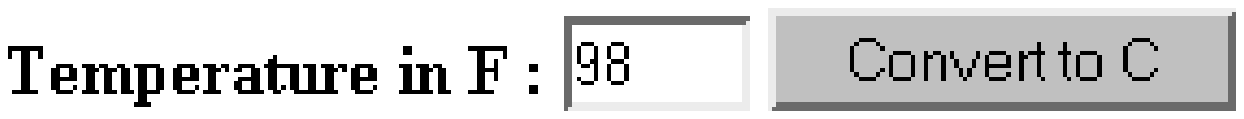}
  \caption{Input page in WWW browser.}
  \label{fig;shtml}
\end{figure}

\begin{figure}[!tb]
  \centering \includegraphics[angle=0,width=2.8in]{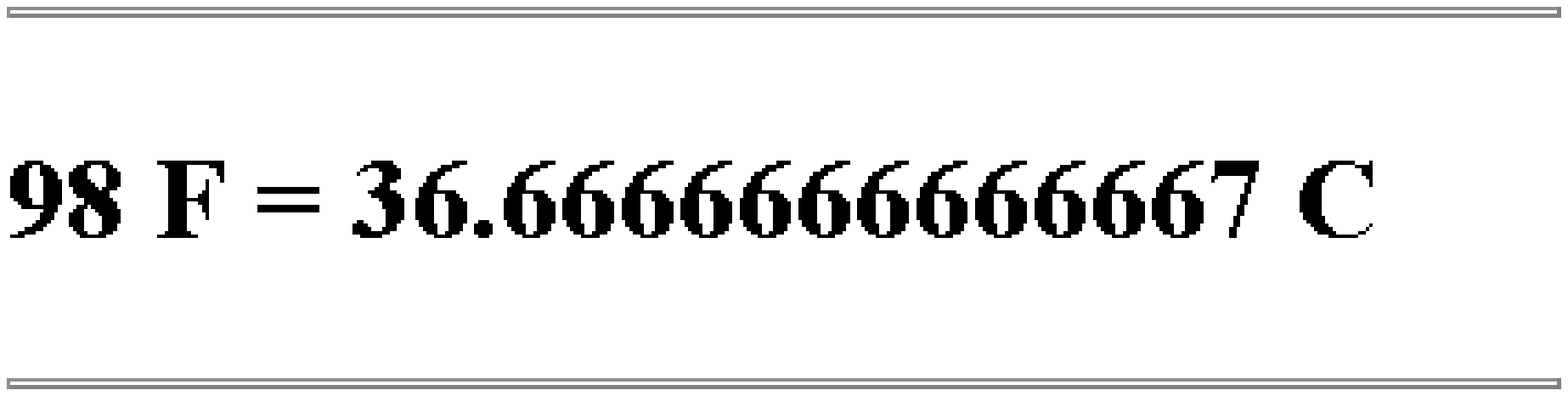}
  \caption{Output of temperature converter}
  \label{fig;scgi}
\end{figure}

As a simple example, we will present a WWW based tool that converts
temperatures from Fahrenheit to Celsius. The conversion is
\begin{equation}
   T_{C} = \frac{5}{9} (T_{F} - 32.0)
   \label{eq;conv}
\end{equation}
The first part is to create the WWW page that contains the form that
allows one to enter the parameters, in our case the temperature in F. The
HTML code for our example page is listed in Fig. \ref{code;shtml}.
\begin{figure}[!tb]
\footnotesize
\begin{verbatim}

<html>
<head>
  <title>Temperature converter</title>
</head>
<body>
<form method="get"
      action="http://localhost/cgi-bin/temp.cgi">
  <b>Temperature in F :</b>
  <input type="text" size="5" name="TEMP">
  <input type="submit" value="Convert to C">
</form>
</body>
</html>
\end{verbatim}
\caption{HTML code for input WWW page of the temperature converter.}
\label{code;shtml}
\end{figure}
After the usual HTML header lines, the form is started with the tag
\texttt{form}. The field \texttt{action} specifies the location of the
script that will process the form input. Obviously one needs to have a WWW
server running that allows one to run scripts in the \texttt{cgi-bin}
directory. The tag \texttt{input} specifies the input field for the
temperature that we want to convert. Note the field \texttt{name="TEMP"}
which labels this input field. This label will be used later in the script
that is called once the \texttt{Convert to C} button is pressed. The
actual appearance of our little form in a WWW browser is shown in Fig.
\ref{fig;shtml}.
Next we need to discuss the script that will actually obtain the value we
entered in the form, execute the temperature conversion and display the
result in the WWW browser. Although one can use a variety of programming
languages for the script in question, a very common and easy to use
language for this type of scripts in PERL. The listing of the compete
script of our temperature converter is listed in Fig. \ref{code;scgi}.
\begin{figure}[!tb]
\footnotesize
\begin{verbatim}

use CGI;

$cgi=CGI::new();
$ftemp=$cgi->param('TEMP');
$ctemp=5.0/9.0*($ftemp-32.0);

print $cgi->header()."\n";
print $cgi->start_html(-title=>'T. converter');
print "<hr><h1>$ftemp F = $ctemp C</h1><hr>\n";
print$cgi->end_html()."\n";
\end{verbatim}
\caption{CGI script of the temperature converter.}
\label{code;scgi}
\end{figure}
In the first two lines the module CGI is loaded and a new CGI object is
initialized. Next the value entered in the form is obtained using the
statement \texttt{\$ftemp=\$cgi->param('TEMP')}. Note that \texttt{TEMP}
is the value of the name field of the corresponding input fiels in the
HTML page listed above (Fig. \ref{code;shtml}). Next we calculate the
corresponding temperature in degrees Celsius using Eq. \ref{eq;conv}. The
last four lines of the script produce the HTML output. First a correct
HTML header is written, followed by the title and the line containing the
result. Finally the HTML document is closed. The output from our
conversion script as is appears in a WWW browser is shown in Fig.
\ref{fig;scgi}.
Hopefully this simple example has illustrated how easy it is to create
interactive WWW pages. As we have mentioned before, a script called by a
WWW form can also execute more complex external programs, allowing one to
use the WWW as a computer platform independent graphical users interface.
Obviously the result can also be in graphical form (see interactive guide
on diffraction physics \cite{prnebi01}). There is also a large number of
books about creating WWW pages.
%
\section{Examples}

In this section we will briefly discuss three scripts that we think are
useful to users. The scripts and more information can be obtained by
contacting the authors.

\subsection{Activate}

Before putting a sample in the beam of an instrument at a spallation
neutron source, it is helpful to estimate how active the sample will get
as a result of being irradiated by neutrons for the duration of the
experiment. Although this estimate is straight forward and in many cases
the needed material constants, are listed in the beamline documentation,
it is still a barrier to the experimentalist who has to worry about many
details of the experiment. To make the activation estimate as easy as
possible, we have developed a WWW interface that allows one to enter
sample composition, sample mass and neutron beam current. As a result the
estimated storage time as well as additional information is listed. The
script also displays an extensive help text that aids users to understand
the resulting numbers and it also emphasizes that fact that is is a simple
\textit{estimate} and not a details activation calculation.

\subsection{Beam time scheduler}
\begin{figure}[!tb]
  \centering \includegraphics[angle=0,width=3.25in]{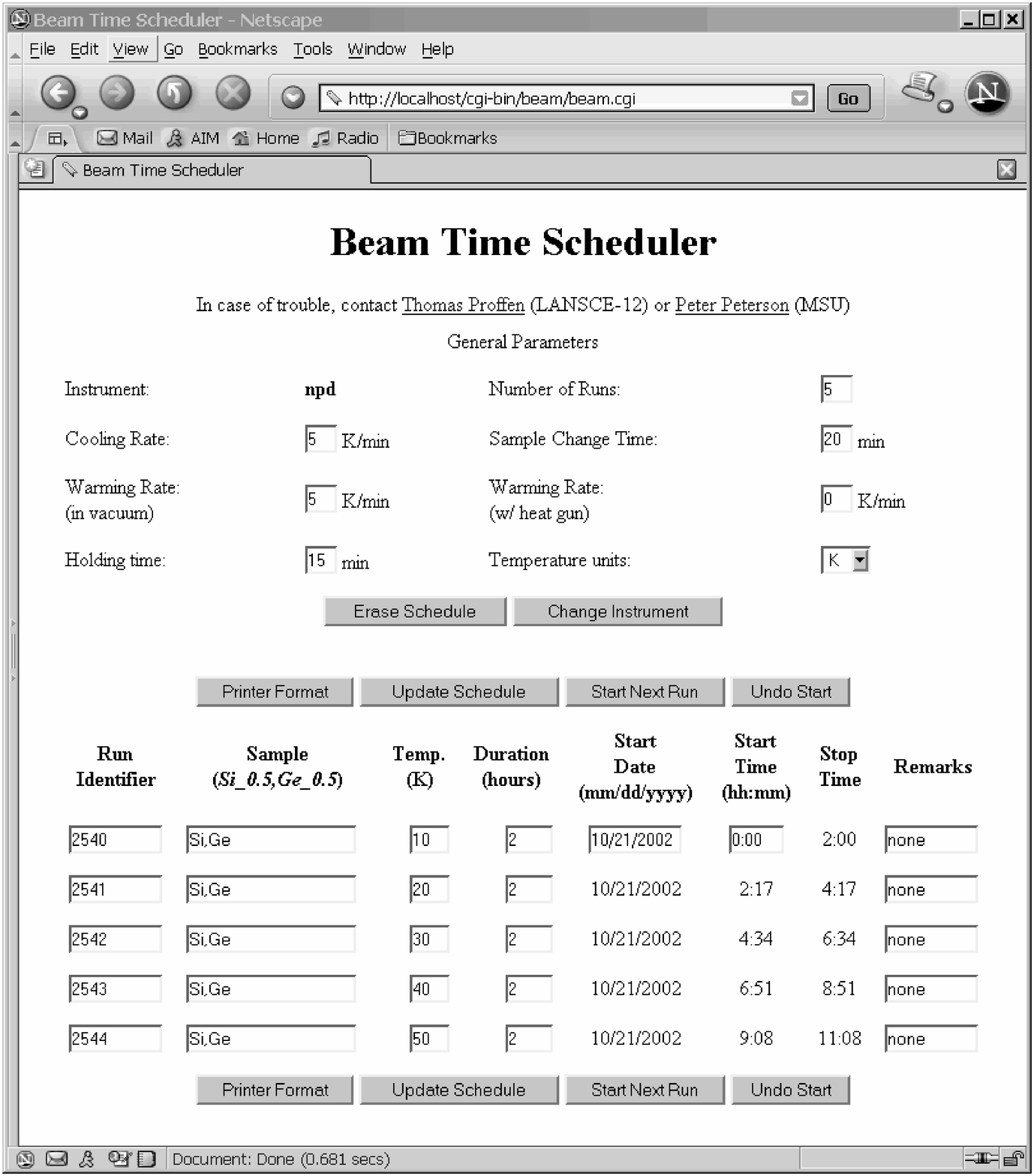}
  \caption{Screen shot of beam time scheduler}
  \label{fig;beam}
\end{figure}
Another common task during an experiment is to estimate a schedule of the
experiment taking into account sample cooling and warming rates, measuring
time and the time required to change a sample. This helps to determine
when the experimenter needs to be back at the instrument to e.g. change
the sample and how many temperature points one can fit in the given beam
time allocation. However, as things change during the experiment, this
schedule needs to be updated and one might need to decide to drop a
temperature point or reduce the measurement time. A WWW based tool that
aids these estimates is the beam time scheduler. A screen shot is shown in
Fig. \ref{fig;beam}. The tool allows one to specify sample cooling and warming
rates as well as the typical sample change time. Then the sample
compositions and measurement times are entered. By comparing temperatures
and sample compositions, the script determines the total time required for
a measurement and calculated starting data and time for each experiment.
Once a run is started, the real starting time is entered and all the
following times are recalculated. This way one has always am as accurate
as possible estimate of the starting times for all subsequent
measurements, assuming of course nothing goes wrong. The authors have used
this tool many times and found it to be a very helpful tool. The printer
format button creates a table suitable for printing and adding to the
experiment log book.

\subsection{Experimental data database}

The last tool available is a database for experimental data. Although
most user facilities have data bases for all experimental data collected,
a single research group might collect data at many different facilities.
The experimental data data base \textit{PDFsearchN} is a simple WWW based
tool to search a local data base of experimental data. The WWW search tool
is accompanied by a check-in and check-out script for those data. The idea
is that after an experiment is done, the data and all relevant experiment
information are entered. This way the information written down in
individual experiment log books become easily available to the research
group and finding data several years after they were measured becomes very
simple.

\section{Conclusions}

The WWW is the perfect way to distribute documentation and interactive
tools in a simple and computer platform independent way. Most people are
familiar with using a WWW browser and having the information "on the net"
also allows users to learn about a particular instrument before they
actually arrive at the facility. Apart from documentation, there are many
examples of helpful tools that can be implemented as interactive WWW
forms. In this paper we have presented a simple example of a temperature
converter, illustrating how easy it is to develop those interactive WWW
based tools. We also have shown three simple tools that might be
particularly helpful to users of large radiation facilities. Comments or
requests for software or more information are welcome by the authors.


\begin{acknowledgments}
Argonne National Laboratory is funded by the U.S. Department of
Energy, BES-Materials Science, under Contract @-31-109-ENG-38. Los
Alamos National Laboratory is funded by the US Department of Energy
under contract W-7405-ENG-36.
\end{acknowledgments}

\bibliography{c:/thomas/documents/bib/thomas/mypub,%
              c:/thomas/documents/bib/thomas/mycabs}
\bibliographystyle{aip}

\end{document}